\documentclass[aps,prd,twocolumn,a4paper,nofootinbib,superscriptaddress]{revtex4-1}

\usepackage{amsmath,amssymb}
\usepackage{hyperref}
\usepackage{bbm}
\usepackage{slashed}
\usepackage{graphicx}
\usepackage{mathptmx}
\usepackage{xcolor}

\usepackage{verbatim}

\newcommand{\bw}{\begin{widetext}}
\newcommand{\ew}{\end{widetext}}

\newcommand{\be}{\begin{equation}}
\newcommand{\ee}{\end{equation}}
\newcommand{\bestar}{\begin{equation*}}
\newcommand{\eestar}{\end{equation*}}

\newcommand{\bi}{\begin{itemize}}
\newcommand{\ei}{\end{itemize}}
\newcommand{\bea}{\begin{eqnarray}}
\newcommand{\eea}{\end{eqnarray}}

\newcommand{\hbo}{\hbox to 1 true cm {\hfill } }

\newcommand{\ud}{\mathrm{d}}

\begin{document}
\title{Laser intensity effects in noncommutative QED}

\author{Thomas Heinzl}\email{theinzl@plymouth.ac.uk}
\affiliation{School of Computing and Mathematics, University of Plymouth, Plymouth PL4 8AA, UK}

\author{Anton Ilderton}\email{anton.ilderton@physics.umu.se}
\affiliation{Department of Physics, Ume\aa\ University, SE-901 87 Ume\aa, Sweden}

\author{Mattias Marklund}\email{mattias.marklund@physics.umu.se}
\affiliation{Department of Physics, Ume\aa\ University, SE-901 87 Ume\aa, Sweden}

\begin{abstract}
\noindent We discuss a two-fold extension of QED assuming the presence of strong external fields provided by an ultra-intense laser and noncommutativity of spacetime. While noncommutative effects leave the electron's intensity induced mass shift unchanged, photons change significantly in character: they acquire a quasi-momentum that is no longer light-like. We study the consequences of this combined noncommutative strong-field effect for the basic lepton-photon interactions.
\end{abstract}

\maketitle

\paragraph*{Introduction.} Within the next few years, laser facilities such as ELI \cite{ELI:2009} will allow access to the uncharted standard model sector of low energy, strong field QED (and possibly beyond), exploring nonlinear vacuum phenomena \cite{Dittrich:2000zu,Marklund:2008gj,Heinzl:2008an}, nonperturbative pair creation \cite{Schwinger:1951nm,Dunne:2008kc}, beamstrahlung in accelerators \cite{Chen:1988ec} and searches for new light particles \cite{Ahlers:2007qf,Gies:2008wv}. Research into these topics has become an extremely active area and is summarised in the reviews \cite{McDonald:1986zz, Mourou:2006, Marklund:2006, Salamin:2006ff}. In this paper we investigate the relevance of ultra-high intensities for physics beyond the standard model not directly related to hitherto unobserved particles. Instead, our chosen model of new physics is the noncommutativity of spacetime, and the formalism noncommutative QED (NCQED for short). Noncommutativity appears in many guises, either via space-time foaminess in quantum gravity schemes \cite{Amelino-Camelia:2005, Lammerzahl:2006} or through noncommutative field theory extensions of the standard model \cite{Douglas:2001ba, Szabo:2001kg}, and may play a role in describing physics somewhere between current (i.e. LHC) and Planck scales. Note that it is not necessary to assume that the noncommutative scale matches the Planck scale (or the string scale) \cite{Abel:2006wj}, but even if it does, noncommutative effects can appear at much lower energies, and have macroscopic effects, through UV/IR mixing \cite{Helling:2007zv, Palma:2009hs}.

Clearly, detection of effects due to noncommutativity would be of profound importance for our understanding of spacetime structure, and therefore possible tests of noncommutativity are of interest to a wide community.  While many astrophysical tests have been put forward (see e.g.\ \cite{Amelino-Camelia:1998}, where dispersive corrections to photon propagation are discussed), such searches remain difficult due to the nature of the obtained data, the collection of which may depend on random astrophysical events (see, though \cite{Palma:2009hs}). As controlled laboratory/accelerator tests \cite{Hewett:2000zp} occupy a complementary regime of momentum space compared to astrophysical tests, they are highly relevant as a dual means of putting limits on modifications to the standard model, as already noted in the context of particle searches \cite{Gies:2008}.

We discuss here two elementary processes of strong-field QED, namely nonlinear Compton scattering and pair production \cite{Nikishov:1963,Nikishov:1964a} in an intense laser beam. Momentum conservation forbids either of these processes from occurring in vacuum, in both QED and NCQED. Strong-field NCQED therefore provides another new arena of physics generated by the interplay of noncommutative and intensity effects. Our aim is to understand the form of some of these effects, present some results with a nonperturbative dependence on the strong field and the noncommutativity tensor, and see if large intensity can in some way compensate for the, typically, low energy of laser experiments in probing new physics.

Intensity effects are parameterised by $a_0$, the dimensionless laser amplitude. It is Lorentz and gauge invariant \cite{Heinzl:2008rh}, and describes the ratio of the electromagnetic energy gain of an electron, across a laser wavelength, to its rest mass. For Petawatt class lasers $a_0$ is of order $10^2$, while $a_0\sim 10^3$--$10^4$ will be achieved with the next generation of facilities. At this point it is important to stress that there is a natural upper limit for $a_0$ beyond which any laser beam will become unstable \cite{Bulanov:2004de}. Recall the common expression relating $a_0$ to the electric field strength, $E$, and frequency $\omega$ of the laser,
\be\label{a0}
	a_0 = \frac{eE}{m\omega} =: \frac{1}{\nu} \frac{E}{E_{\text{crit}}} \; ,
\ee
where $m$ is the electron mass, $\nu = \omega/m$ and $E_\text{crit} = m^2/e$ is Schwinger's critical field strength \cite{Schwinger:1951nm}. If the electric field exceeds this value, photons will transmute into pairs as the vacuum is ``boiled" \cite{Ringwald:2003iv}, and the beam will be destabilised. This implies an upper bound on $a_0$: for $E = E_\text{crit}$, (\ref{a0}) becomes $a_{0,\mathrm{max}} = 1/\nu \simeq 5\times 10^5$ for optical lasers. We are therefore prevented from taking intensity to be arbitrarily large.

\paragraph*{Strong-field QED.} As an illustration consider nonlinear Compton scattering \cite{Harvey:2009ry}, specifically the head-on collision of an electron (momentum $p$) with an intense laser, and the scattering of a photon (momentum $k'$) out of the beam. Photon numbers in an intense laser are sufficiently high for them to be treated collectively as a classical background field, which we take to be a circularly polarised plane wave
\be\label{circ}
	a^\mu(k.x) = a_1^\mu\cos(k.x) + a_2^\mu\sin(k.x)\;,
\ee
with $k^2 = 0 = a_j.k$ and $a_i.a_j = -a^2 \delta_{ij} \leq
0$. The interactions of this field with (both scattered and
virtual) quantum particles are included via a dressing, by
the external field, of all fermion lines in Feynman
diagrams; internal lines (free propagators) are replaced by
Volkov propagators, while external lines (free spinor
wavefunctions) are replaced by Volkov wavefunctions
\cite{Volkov:1935}. The only interaction between quantised
fields is that of ordinary QED, and at tree level the
associated S-matrix element is given by the Feynman diagram
\be\label{3pt}
	\parbox{0.7cm}{$S_\text{fi} :=$}
\parbox{3cm}{\includegraphics[width=0.25\columnwidth]{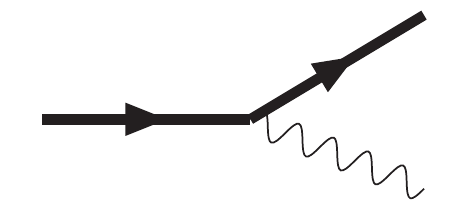}}
\ee
where heavy lines represent Volkov electrons. $S_\text{fi}$ can be calculated exactly in $a^\mu$, which is important
since high intensity backgrounds are nonperturbative in $a_0$. Notably, we will see that an exact treatment is also
possible in NCQED. $S_\text{fi}$ takes the form of an infinite sum of contributions supported on the delta
functions $ \delta^4 \left( q_\mu + n k _\mu  - q'_\mu - k'_\mu \right) $ for $n \in [1,\infty)$. Here, $q$ is the
quasi-momentum
\be \label{quasi.mom}
	q_\mu = p_\mu + a_0^2m^2 k_\mu/2k.p\;,
\ee
(similarly, for the scattered electron, $p'\to q'$) with intensity parameter $a_0 = ea/m$ for the field (\ref{circ}). Squaring, we find
\be \label{mass.shift}
  q^2 = m^2(1+a_0^2) =: m_*^2 \; ,
\ee
which is the intensity dependent mass shift of an electron in a laser background \cite{Sengupta:1952}. Hence, each
contribution to the cross section corresponds to an effective process in which a heavy electron, of mass $m_*$,
interacts with $n$ laser photons of momentum $k_\mu$ and emits a scattered photon of momentum $k'_\mu$, i.e.\ $e^-(q)
+ n\,\gamma(k) \to e^-(q') + \gamma(k')$. These multi-photon subprocesses are the origin of the name
``nonlinear'' Compton scattering. Despite not being described in terms of genuine on-shell momenta,
contributions from higher harmonics ($n>1$) are experimentally distinguishable \cite{Chen:1998}. The basic
intensity effect is the mass shift (\ref{mass.shift}). As the electron ``gains weight'' it recoils less, which reduces
the energy transfer to the emitted photon. As a result, for an electron with gamma factor $\gamma$, there is an overall redshift of the photon emission spectrum with the $n=1$ Compton edge changing from $\nu' \simeq 4 \gamma^2 \nu$ to $4 \gamma^2 \nu/ (1 + a_0^2 + 4\gamma\nu)$, for $\nu'$ the  scattered photon frequency in units of $m$, and assuming typical parameter values. Further details may be found in \cite{Harvey:2009ry} where the resulting spectra are discussed extensively. Experiments to measure
these spectra are planned at the facilities COBALD (Daresbury Labs) and DRACO (Forschungszentrum
Dresden-Rossendorf).

\paragraph*{Strong-field NCQED.} We take constant noncommutativity $[x^\mu,x^\nu] = i \Theta^{\mu\nu}$,
generated by the usual Moyal star product, $f(x) \star g(x) = f(x)\, \exp \big[ i \partial_L \wedge \partial_R \big] \,
g(x)$ with $\partial_L$ ($\partial_R$) acting to the left (right) and $a \wedge b := \tfrac{1}{2} a_\mu
\Theta^{\mu\nu} b_\nu$. We also write $|\Theta| \sim 1/\Lambda^2$ with $\Lambda$ the noncommutativity scale. The
calculation of amplitudes in NCQED is discussed thoroughly in \cite{Hayakawa}. The inclusion of a background field,
i.e.\ turning on $a_0 > 0$, affects the fermion sector in a manner similar to that described above; free fermion
propagators become dressed, and electron wavefunctions become modified Volkov wavefunctions, $\Psi_p = \exp (iS[\tilde{a}]) \Gamma [\tilde{a}] u_p$ \cite{AlvarezGaume:2000bv}, where $S$ is the Hamilton-Jacobi action for an electron in a plane wave field $\tilde{a}$, $\Gamma[\tilde{a}]$ a field dependent combination of Dirac matrices and $u_p$ a free Dirac spinor. Compared to the commutative case the electron sees a {\it different} background field $\tilde{a}^\mu(x) := a^\mu(k.x + k \wedge p)$, the phase $k.x$ shifted by the noncommutative contribution $k \wedge p$. However, upon averaging the \textit{periodic} field (\ref{circ}) over one laser wavelength, we have $\langle{\tilde a}^2 \rangle = \langle {a}^2 \rangle $: we conclude that, for periodic fields, the electron mass shift in NCQED is the {\it same as in QED}, namely (\ref{mass.shift}). This should not be expected to hold for finite beam geometries such as pulses, though.

Things are more interesting for the photons. As the
noncommutative U(1) gauge group is non-abelian, the photon
is self interacting in NCQED. An immediate consequence is
that photons see the background field and become dressed by
it, as the fermions do. Working in background covariant gauge, external
photon lines change from $\exp(-ik'.x)\epsilon^\mu$ in QED
(with polarisation vector $\epsilon^\mu$) to
\bestar
	\exp\bigg(-ik'.x -\frac{i}{2k.k'}\int\limits^{k.x}\! 2e {\mathbb A}.k'-e^2\mathbb{A}^2\bigg)\exp\bigg[\frac{e}{k.k'}k^{[\mu} {\mathbb A}^{\nu]}\bigg]\epsilon_\nu \;,
\eestar
with $k'.k'= k'.\epsilon=0$. The matrix exponential truncates at the second order. The background seen by the photon is $\mathbb{A}^\mu(k.x) := a^\mu(k.x + k\wedge k') - a^\mu(k.x - k\wedge k') $ which illustrates the extended nature of the photon in NCQED; it can be interpreted as a ``dipole'' of length $2k\wedge k'$ with charges at its endpoints, as previously observed for adjoint matter \cite{AlvarezGaume:2000bv}. Note that the larger the momentum $k'$, the longer the dipole -- this is a simple manifestation of the UV/IR mixing inherent to noncommutative theories. Further discussion of this solution will appear in \cite{sigma}.

We are now ready to present S-matrix elements for our NCQED processes. At tree level the only contributing diagram is the analogue of (\ref{3pt}), with vertex $\bar\psi \star \slashed{A} \star \psi$.

\paragraph*{Nonlinear Compton scattering.} Mod-squared and summed over polarisations, the S-matrix element is again an
infinite sum over harmonic processes,
\be\label{S}
	\frac{1}{VT}\sum\limits_\text{pols}|S_\text{fi}|^2=\sum\limits_{n=1}^\infty\, \mathcal{S}_n\, \delta^4 \left( q_\mu + n k_\mu  - q'_\mu - l'_\mu\right)\;.
\ee
The support of the delta functions is almost as in the commutative case; $q$ and $q'$ are as in (\ref{quasi.mom}). Noncommutative input to the kinematics comes from the photon momentum $l'$, which is related to the asymptotic photon momentum $k'$ by
\be \label{l.prime}
	l'_\mu = k'_\mu + \frac{2a_0^2m^2\sin^2(k\wedge k')}{k.k'}k_\mu \; ,
\ee
and, like the electron quasi-momentum (\ref{quasi.mom}),
features a longitudinal piece along the laser momentum $k$.
Notably, $l'$ no longer squares to zero, but rather
\be
  l^{\prime 2} = 4 m^2 a_0^2 \sin^2(k\wedge k') =: 4 m^2 \Delta^2 \; .
\ee
This may be interpreted as a field and momentum dependent
``photon mass'' (or self-energy) and encodes the
modification to probe photon propagation from interactions
with the background field. It is naturally associated with a
photon length scale $\lambdabar_l := 1/2m\Delta$, which
again points towards an effective finite extension of the
noncommutative photon.

Introducing the variable $x:=k.k'/k.p'$, the constants $y_n:= 2n k.p/m_*^2$ and the function
\be\label{z}
	z^2 := \frac{4m^2}{y^2_1 m_*^2} \bigg( xy_n-x^2-\frac{4\Delta^2(1+x)}{(1+a_0^2)} \bigg) \bigg[ a^2_0 + 4 \frac{\Delta^2(1+x)}{x^2} \bigg]\;,
\ee
the amplitudes in (\ref{S}) take the form
\be\label{ampsq}\begin{split}
  \mathcal{S}_n = -2J^2_n + a_0^2\bigg(1+\frac{x^2}{2(1+x)}\bigg) \big( J^2_{n+1} + J^2_{n-1}
- 2J^2_n \big) \\
  + 2\Delta^2\bigg(1+2\frac{1+x}{x^2}\bigg)\bigg[J^2_{n+1} + J^2_{n-1} -2J^2_n\bigg] \;,
\end{split}\ee
where all Bessel functions $J$ have argument $z$ from (\ref{z}). These results are exact in both the noncommutativity parameter and in the background field. The second line in (\ref{ampsq}) describes purely noncommutative effects. In the limit $\Delta\to 0$ (\ref{z}) and (\ref{ampsq}) recover the strong-field QED results of \cite{Narozhnyi:1964}.

It is important to note that {\it all} noncommutative effects discussed in this paper are generated by (the square of) the photon mass, $2m \Delta$. Let us now discuss the form of these effects, and how their size depends on energy and intensity. NCQED cross sections acquire new angular dependencies, relative to their QED limits, due to the breaking of Lorentz invariance caused by the preferred directions inherent in $\Theta^{\mu\nu}$. These appear through the combination $k \wedge k' \equiv k_-\Theta^{-\nu}k'_\nu$, so our laser experiment is
specifically sensitive to the background-induced four-vector $N^\nu = \Theta^{-\nu}$, i.e. to {\it lightlike noncommutativity} \cite{Aharony:2000gz}.
In a head-on collision of electrons and a circularly polarised laser, the presence of the preferred direction $N^\nu$ breaks the azimuthal symmetry around the beam axis. This causes an oscillatory dependence on the azimuthal angle $\phi$ in the partially integrated cross section, $\sigma' := \ud\sigma/\ud\phi$, which in QED is $\phi$ independent. A second interesting effect is broadening of spectral lines: in QED a line spectrum appears for special intensity values defining effective centre-of-mass frames \cite{Harvey:2009ry}. For instance, if $a_0 \simeq 2 \gamma$, the spectral range of the fundamental harmonic shrinks to a line located at $\nu' = \nu$. In NCQED, however, this spectral line acquires a $\phi$ dependent finite width, $(\Delta^2/\nu)\cos^2 \phi$. Hence, spectral gaps are reduced in NCQED.

We now turn to the size of such effects. To this end we approximate
\be\label{Deltasimp}
	\Delta \simeq a_0 \,k \wedge k' \simeq a_0 \nu \nu' m^2/\Lambda^2\;,
\ee
suppressing angular dependencies in the last identity. At first glance, it seems that the smallness of $k \wedge k'$ can be compensated to an arbitrary degree by increasing the intensity $a_0$. However, this is misleading, both because there is an upper limit to $a_0$, as discussed, and because, in this process, $\nu'$ is a function of $a_0$. Adopting the strong-field QED value for $\nu'$ (noncommutative corrections are higher order in $\Delta$) we find for the backscattered fundamental harmonic ($n=1$),
\be \label{delta.estimate}
  \Delta \simeq 4 \gamma^2 \nu^2 \frac{a_0}{1 + a_0^2 + 4\gamma \nu} \frac{m^2}{\Lambda^2} \; .
\ee
For optical lasers we have $\nu\simeq 2\times 10^{-6}$ while at a future linear collider (say ILC or CLIC) one may reach electron energies of  a few TeV ($\gamma = 10^7$) such that $\gamma\nu \simeq 10$. Thus, most of the enhancement in (\ref{delta.estimate}) has to come from the second, $a_0$ dependent term. This becomes maximal for $a_0 = (1 + 4\gamma \nu)^{1/2}$, so we obtain the order-of-magnitude estimate  $\Delta \simeq 10 m^2/\Lambda^2$.
The intensity, although large, is unable to enhance the noncommutative signal in this process because $k'_\mu$ is determined by the scattering kinematics, and itself deviates only slightly from QED. To increase the size of the noncommutative signal we need to enhance $\Delta$, which suggests looking for a process where both laser and probe photon momenta are {\it incoming} and hence {\it independently} tuneable in magnitude. Such a process is multi-photon pair production.
	
\paragraph*{Pair production.} Strong-field pair creation is obtained from (\ref{3pt}) via crossing symmetry and was first observed in the SLAC E-144 experiment about a decade ago, where a Terawatt laser was brought into collision with the 50 GeV SLAC beam \cite{Burke:1997ew,Bamber:1999zt}. We consider now the noncommutative pair production processes $n\gamma + \gamma' \to e^- + e^+$ where $n$ laser photons collide with a high-energy photon $\gamma'$. Momentum conservation for the different harmonic processes, cf.~(\ref{S}), becomes $n k + l' = q + q'$ with $l'$, $q$ and $q'$ now being momenta for the incoming photon, outgoing electron and outgoing positron, respectively, defined just as in (\ref{quasi.mom}) and (\ref{l.prime}). Unlike Compton scattering, strong-field pair creation is a threshold process which requires a minimum energy input or, equivalently, a minimum number of laser photons, which in QED is $n_0 := 2m_*^2/k.k'$ \cite{Bamber:1999zt}. Following \cite{Narozhnyi:1964}, we introduce the variable $u$ and constants $u_n$ defined by
\be
	u := \frac{(k.k')^2}{4k.p\ k.p'}\;, \qquad u_n := \frac{n}{n_0}\;.
\ee
The variable (\ref{z}) is replaced by
\be
	z^2 := \frac{4u(u_{n,\Delta}-u)}{u^2_1(1+a_0^2)}\bigg[a^2_0
-\frac{\Delta^2}{u}\bigg]\;,  \quad
u_{n,\Delta}:=u_n+\frac{\Delta^2}{1+a_0^2}\;,
\ee
and the scattering amplitudes (obtained from (\ref{ampsq}) by
crossing symmetry \cite{Narozhnyi:1964}) are now
\be\label{ampsPP}\begin{split}
	\mathcal{S}_n = 2J^2_n + a_0^2 \big(2u-1\big) \big(
J^2_{n+1}+J^2_{n-1}-2J^2_n \big) \\
	 -
2\Delta^2\bigg(1-\frac{1}{2u}\bigg)\big(J^2_{n+1
}+J^2_{n-1}-2J_n^2\big) \;,
\end{split}
\ee
again suppressing the common argument $z$ of the Bessel functions. As $\Delta \to 0$ the first line, and $z$, recover the QED results given in \cite{Narozhnyi:1964}. The second line in (\ref{ampsPP}) again describes purely noncommutative effects.

Azimuthal dependencies in the cross section may be seen in \textit{transverse} collisions between the (circularly polarised) laser and probe photon ($\theta = \pi/2$). As the probe's azimuthal angle of approach, $\phi$, changes so will the cross section, due to the presence of $N^\nu$. The pair production rate, unlike that in QED, will therefore exhibit a periodic dependence on the collision angle. The noncommutative cross section is actually \textit{larger} than its QED counterpart, and we have checked that imperfect circular polarisation of the beam causes oscillations \textit{around} the constant QED average, and thus can be clearly distinguished in shape from the noncommutative effect. A more novel effect is seen in the noncommutative counterpart of the photon number threshold $n_0$, which is
\be \label{threshold}
  n_{0, \Delta} := {2m^2(1 + a_0^2 - \Delta^2)}/{k.k'} < n_0 \; .
\ee
We note that the combination of noncommutativity and strong fields reduces the pair production threshold by an amount proportional to the ``photon mass'' squared. The reason is that the probe photon itself drags additional energy from the background into the collision, in the form of its mass, or dressing, so fewer additional photons are required to reach a particular threshold. Not unexpectedly, it turns out that $\Delta \ll a_0$ for realistic noncommutativity scales and beam
parameter values. Nevertheless, as the incoming asymptotic momenta, $k$ and $k'$, are now independent we can significantly increase $\Delta$ relative to the nonlinear Compton result, by making the associated (lab) energies, $\nu$, $\nu'$ \textit{and} the intensity parameter $a_0$ as large as possible.

Let us now discuss the size of noncommutative effects in pair creation, and bounds on the noncommutativity parameter which may be obtained at laser facilities.  We assume a scenario similar to the SLAC E-144 experiment \cite{Burke:1997ew,Bamber:1999zt}, namely that probe photons are generated via Compton backscattering of {\it low intensity} photon beams off high energy electron beams as will be available at the ILC or CLIC, so $\gamma=10^7$. These parameters put us in the ``linear" regime in terms of photon frequency boosting, meaning we can obtain $\nu' = \gamma=10^7$ from the (inverse) Compton blue shift \cite{McDonald:1986zz}. These backscattered photons are then collided with a high intensity optical laser ($a_{0} \gg 1$) to produce pairs. Adopting the ELI value $a_0 \simeq 10^3$, one finds $\Delta \simeq 10^{4} m^2/\Lambda^2$. Adopting instead the theoretical maximum intensity such that $a_0 \nu = 1$, we have  $\Delta \simeq 10^{7} m^2/\Lambda^2$, exceeding the nonlinear Compton value by 6 orders of magnitude.

Following the example of \cite{Carroll:2001ws}, we can also use these results to derive simple bounds on the noncommutativity scale. To do so we need an estimate for the percentage error in our measurements of the equivalent commutative processes. There is little detailed data available from the experiments which have investigated pair production and nonlinear Compton, although SLAC and other experiments are in agreement with QED, in particular as regards the mass shift of the electron \cite{Bamber:1999zt, Meyer, Moore:1995zz} and higher harmonic generation \cite{Chen:1998}. Much more data will appear in the near future following experiments at COBALD, DRACO and, a little later, Vulcan 10PW. Here, then, we will give an estimate for the bounds which could be obtained assuming that the QED predictions for these nonlinear vacuum phenomena continue to agree with theory. Suppose we can obtain an experimental sensitivity of $P\,\%$. The noncommutative corrections to our processes are controlled by $\Delta^2$,  so it follow from (\ref{Deltasimp}) that we can read off a bound of
\be
	|\Theta^{-\perp}|< \big(10^3 P^{-\tfrac{1}{4}}(5a_0\nu')^{\tfrac{1}{2}}\text{eV}\big)^{-2}\;,
\ee
where we have used optical frequency for the laser. To illustrate, we take $P=1$. For $\gamma$ as above and ELI intensity, the bound becomes
\be
	|\Theta^{-\perp}|<(0.2\text{ GeV})^{-2}\;,
\ee
while for the theoretical limit, $a_0 = 5\times10^5$, where the intensity compensates maximally for the low energy of the laser, we find
\be
	|\Theta^{-\perp}|<(5\text{ GeV})^{-2}\;.
\ee
Note again that these bounds constrain the \textit{light-like} entries $\Theta^{-\perp}$ and thus complement the results obtained in \cite{Carroll:2001ws}.
\paragraph*{Conclusions.} We have studied noncommutative QED in the presence of ultra-intense laser fields. Our main findings are an exact form for the ``dressed photon" in noncommutative QED, similar to the Volkov electron and, as a consequence, an effective ``photon mass'', $2m \Delta$, depending both on intensity and noncommutativity parameters. Typical processes such as Compton scattering and pair creation can be calculated exactly in the background field, and acquire corrections of order $\Delta^2$. Even if possibly too small to be observable they may still be used, as we have discussed, to obtain limits on specific entries of the noncommutativity tensor, in particular its \textit{lightlike} components. It would be interesting to apply our approach to the study of noncommutative corrections to vacuum birefringence \cite{Abel:2006wj}. 

M.~M.\ and A.~I.\ are supported by the European Research Council under Contract No.\ 204059-QPQV, and the Swedish Research Council under Contract No.\ 2007-4422.

\end{document}